\documentclass[prl,twocolumn]{revtex4-2}
\usepackage{times}
\usepackage{amsmath}
\usepackage{amsfonts}
\usepackage{graphicx}
\usepackage{hyperref}
\usepackage{dcolumn}
\usepackage{bm}

\def\beq{\begin{equation}}
\def\eeq{\end{equation}}
\def\bea{\begin{eqnarray}}
\def\eea{\end{eqnarray}}
\def\beqn{\begin{eqnarray}} 
\def\eeqn{\end{eqnarray}}
\def\beeq{\begin{eqnarray}}
\def\eeeq{\end{eqnarray}}

\def\nn{\nonumber}
\def\Eq#1{Eq.~(\ref{#1})}

\def\lnn#1{\mathrm{log^2}\left(#1\right)}

\def\qon#1{q_{#1,0}^{(+)}}

\def\res#1{{\rm Res}\left(#1\right)}

\def\qb{\mathbf{q}}

\def\lb{\boldsymbol{\ell}}

\def\ii{\imath 0}

\def\res#1{{\rm Res} \left(#1\right)}
\def\ad#1{{\cal A}_{\rm D}^{(#1)}}

\def\ps#1{\widetilde \Delta_{#1}}

\usepackage{xcolor}

\definecolor{darkorange}{rgb}{1.0, 0.55, 0.0}

\begin{document}


\title{On the analytic continuation and matching of threshold singularities from  Vacuum Amplitudes}

\author{Selomit Ram\'{\i}rez-Uribe~$^{(a)}$} \email{selomitru@uas.edu.mx} 
\author{Prasanna K. Dhani~$^{(b)}$} \email{prasanna.dhani@physik.uzh.ch}
\author{German F.R. Sborlini~$^{(c)}$} \email{german.sborlini@usal.es} 
\author{Germ\'an Rodrigo~$^{(d)}$}\email{german.rodrigo@csic.es}
\affiliation{${}^{a}$ Facultad de Ciencias F\'{\i}sico-Matem\'aticas, Universidad Aut\'onoma de Sinaloa, Ciudad Universitaria, CP 80000 Culiac\'an, Mexico. \\
${}^{b}$Physik Institut, Universit\"at Z\"urich, Winterthurerstrasse 190, CH-8057 Z\"urich, Switzerland. \\
${}^{c}$ Departamento de F\'isica Fundamental e IUFFyM, Universidad de Salamanca, 37008 Salamanca, Spain. \\
${}^{d}$ Instituto de F\'{\i}sica Corpuscular, Universitat de Val\`{e}ncia -- Consejo Superior de Investigaciones Cient\'{\i}ficas, Parc Cient\'{\i}fic, E-46980 Paterna, Valencia, Spain.}

\date{\today}

\begin{abstract}
We provide a detailed discussion for all kinematical configurations of the analytic continuation to negative values of initial-state on-shell energies used to generate interferences of scattering amplitudes from vacuum amplitudes in the Loop-Tree Duality. We also extend previous discussions on the matching of threshold singularities and introduce angular averaging as an efficient strategy to achieve a local cancellation when the matching of threshold singularities from vacuum amplitudes is constrained by flavour.
\end{abstract}

\maketitle

\section{Analytic continuation}

In a recent paper~\cite{Ramirez-Uribe:2024rjg}, we have proposed vacuum amplitudes, i.e. scattering amplitudes without external particles, in the Loop-Tree Duality (LTD) representation, as the optimal building blocks to assemble theoretical predictions at high-energy colliders. Explicit examples, such as the decay rate of the Higgs boson to heavy quarks at second order in perturbation theory, have been presented in Ref.~\cite{LTD:2024yrb}. The good behaviour of the integrand has also enabled us to implement and integrate the integrand expression obtained in a quantum computing approach~\cite{deLejarza:2024scm}.

The key observation of Ref.~\cite{Ramirez-Uribe:2024rjg} is based on the fact that the integrand of a multiloop vacuum amplitude in LTD, $\ad{\Lambda}$ with $\Lambda$ the number of loops, is a function of the on-shell energies of the internal momenta $\qon{i_s} = \sqrt{\qb_{i_s}^2+m_{i_s}^2-\ii}$, 
where $\qb_{i_s}$ are the spatial components, $m_{i_s}$ are the internal masses, and the infinitesimal complex prescription $-\ii$ stems from the complex prescription of the original Feynman propagators. Specifically, the only kind of denominators that appear in a vacuum amplitude in LTD, termed causal for the reason explained below, are of the form 
\beq
\ad{\Lambda} \propto \frac{1}{\lambda_{i_1 \cdots i_n}}~,
\label{eq:causal}
\eeq
where $\lambda_{i_1 \cdots i_n} = \sum_{s=1}^n \qon{i_s}$. In general, the LTD representation of a vacuum amplitude will contain different combinations of causal denominators, and numerators (see e.g. Supplemental Material in Ref.~\cite{Ramirez-Uribe:2024rjg}). 

\begin{figure}[t]
\begin{center}
\includegraphics[width=\linewidth]{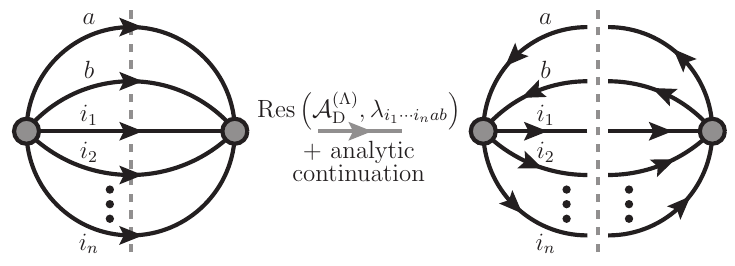}
\caption{Multiloop vacuum configuration encoded by the causal propagator $1/\lambda_{i_1\cdots i_n ab}$. After taking the residue at $\lambda_{i_1\cdots i_n ab} \to 0$ and analytically continuing the on-shell energies of the initial-state particles to negative values, equivalently reversing their momentum flows, we obtain the interference of two scattering amplitudes where all particles along the dividing line are on the shell. \label{fig:analytic}}
\end{center}
\end{figure}

The physical interpretation of \Eq{eq:causal} is particularly interesting. Each causal denominator corresponds to a set of internal propagators that, if connected by a line, would separate the vacuum amplitude into two subamplitudes (see Fig.~\ref{fig:analytic}). Moreover, across this line, all momentum flows are aligned in the same direction. This alignment guarantees that only acyclic configurations are considered, thereby precluding cycles that would break causality. In addition, at $\lambda_{i_1 \cdots i_n}=0$ all the internal momenta of the set would be forced to be on the mass shell. Specifically for vacuum amplitudes, the latter can occur only if $\qon{i_s} = 0$ for all in the set because the on-shell energies are defined as positive. 

In light of these special properties of vacuum amplitudes, the idea introduced for the first time in Ref.~\cite{Ramirez-Uribe:2024rjg} is that the squares and interferences of scattering amplitudes with final-state particles can be generated as residues, called phase-space residues, on the causal denominators
\beq
{\rm Res} \left( \ad{\Lambda}, \lambda_{i_1 \cdots i_n} \right)~,
\eeq
provided that some on-shell energies are analytically continued to negative values. Our original document~\cite{Ramirez-Uribe:2024rjg} clearly says ``{\it by analytically continuing the on-shell energies of the particles that will be identified as incoming to negative values}". This analytic continuation implicitly implies that the momentum flows of the initial-state particles are reversed (see Fig.~\ref{fig:analytic}).

Since the residues and the analytic continuation commute, there are, in fact, three ways to implement this idea: implicit, explicit and semi-implicit. But obviously, all three yield identical results. In the following, $a$ and $b$ label the initial-state particles. 

In the implicit approach, one directly evaluates the residues of the vacuum amplitude at $\lambda_{i_1\cdots i_n ab} = \sum_{s=1}^n \qon{i_s} + \qon{a} + \qon{b} \to 0$ without further modifications. However, in the expression obtained one has to identify the on-shell energies of the incoming particles as negative, i.e., in the center of mass frame, $\qon{a} = \qon{b} =-\sqrt{s}/2$. This approach is very similar to that of many scattering amplitude calculations where one assumes all external particles either outgoing or incoming, but some of the energies are considered negative for specific applications. 

In the explicit approach, one first flips the signs of the on-shell energies of the incoming particles: $\qon{a} \to - \qon{a}$ and $\qon{b} \to -\qon{b}$. The sign flip involves the full expression of the vacuum amplitude, including numerators and denominators. Then, one evaluates the residues of the vacuum amplitude at $\lambda_{i_1\cdots i_n \bar a \bar b} \to 0$ (our convention is to use the bar to indicate a minus sign, i.e., $\lambda_{i_1\cdots i_n \bar a \bar b} = \sum_{s=1}^n \qon{i_s} - \qon{a} - \qon{b} \to 0$). In that case, $\qon{a} = \qon{b} = \sqrt{s}/2$ in the final expression. 

In the semi-implicit approach, one considers that the evaluation of the residue at $\lambda_{i_1\cdots i_n ab}$ usually involves the substitution $\qon{a} \to - \lambda_{i_1\cdots i_n b} < 0$. So, one of the on-shell energies of the incoming particles is implicitly implemented with a negative value. This is the preferred approach for decay rates as the absence of a second incoming particle eliminates the need for additional operations, i.e. sign flips. For scattering processes, however, one can either impose $\qon{b} = -\sqrt{s}/2$, or introduce the sign flip $\qon{b} \to -\qon{b}$ with $\qon{b} = \sqrt{s}/2$.

\begin{figure}[t]
\begin{center}
\includegraphics[width=.9\linewidth]{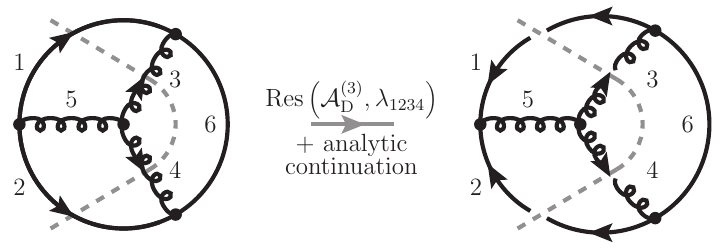}
\caption{The interference of the $s$ and $t$ channels for $q\bar q \to gg$ from the three-loop vacuum amplitude. \label{fig:stchannel}}
\end{center}
\end{figure}

Regarding an explicit calculation, we consider the $s$- and $t$-channel interference for the process $q\bar q \to gg$ (Fig.~\ref{fig:stchannel}), and make the following four-momentum assignments in the vacuum amplitude:
\begin{equation}
q(q_1)~, \quad \bar q(q_2)~, \quad g(q_3)~, \quad g(q_4)~.
\end{equation}
The four-momenta of the other internal propagators are $q_5 = q_1+q_2$ and $q_6 = q_1-q_3$. If ${\cal A}_{\rm D}^{(3)}$ is the corresponding three-loop vacuum amplitude in LTD, the phase-space residue for $\lambda_{1234} \to 0$ gives,
\begin{equation}
{\rm Res} \left( {\cal A}_{\rm D}^{(3)}, \lambda_{1234} \right) \propto \frac{\lambda_{246}\lambda_{24\bar 6}}{\lambda_{345}\lambda_{34\bar 5}}~, \quad {\rm if} 
\quad \lambda_2 = - \sqrt{s}/2~,
\end{equation}
or 
\begin{equation}
{\rm Res} \left( {\cal A}_{\rm D}^{(3)}, \lambda_{1234} \right) \propto \frac{\lambda_{\bar 246}\lambda_{\bar 24\bar 6}}{\lambda_{345}\lambda_{34\bar 5}}~, \quad {\rm if} 
\quad \lambda_2 = \sqrt{s}/2~.
\end{equation}
In both cases the result agrees with the standard calculation, which is $\propto t/s$.

As a final remark, the statement “{\it analytic continuation to negative values}” is unequivocally clear and leaves no room for missinterpretation. Although different conventions can be used to implement this idea, there is no ambiguity whatsoever in the practical encoding of physical processes.

\section{Matching of Threshold singularities}
\label{sec:squared}

\begin{figure}[t]
\begin{center}
\includegraphics[width=\linewidth]{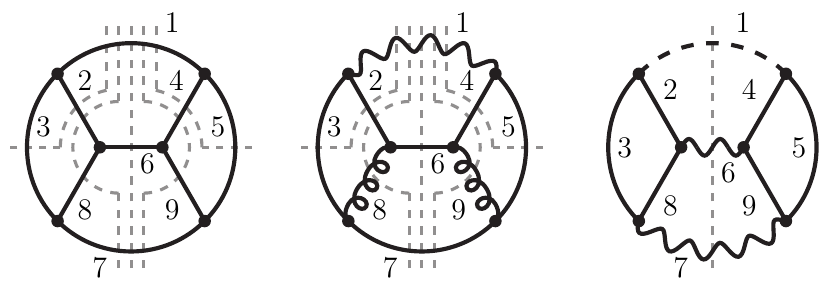}
\caption{Representative four-loop vacuum diagrams contributing at NNLO to the decay rate of a massive scalar particle to scalar particles~(left), $\gamma^*\to q\bar q (gg)$ (middle) and $H\to \gamma\gamma$ (right). The gray dashed lines represent phase-space residues. Only phase-space residues involving loops are represented.
\label{fig:nnlo}}
\end{center}
\end{figure}

In Ref.~\cite{Ramirez-Uribe:2024rjg}, we also demonstrated how threshold singularities match across different phase-space residues, which may lead to a local cancellation of this type of singularity. This local cancellation was made explicit in Ref.~\cite{LTD:2024yrb} for physical processes at next-to-leading order (NLO), and was discussed for a toy scalar process at next-to-next-to leading order (NNLO). Here, we showcase the matching of threshold singularities in more involved cases, in particular considering overlapping threshold singularities. Specifically, we consider the four-loop vacuum diagrams depicted in Fig.~\ref{fig:nnlo}. The internal momenta are defined as 
\bea
&& q_1 = \ell_{3}~,  \qquad q_2 = \ell_{1}~,   \qquad  q_3 = \ell_{13}~, \nn  \\ 
&& q_4 = \ell_2~,    \qquad q_5 = \ell_{2\bar 3}~, \qquad  q_6 = \ell_4~, \nn \\ 
&& q_7 = \ell_{34}~, \qquad q_8 = \ell_{1\bar 4}~, \qquad  q_9 = \ell_{24}~,
\eea
in terms of the four independent loop momenta $\{\ell_s\}_{s=1}^4$. We use the shorthand notation~$\ell_{ij} = \ell_i+\ell_j$ and~$\ell_{i\bar j} = \ell_i-\ell_j$. Nonetheless, the LTD representation in terms of causal propagators and on-shell energies remains unaffected by the choice of the specific loop routing. We will consider $q_1$ as the four-momentum of the initial-state particle. 

The phase-space residues with two particles in the final state feature three different threshold singularities, at $\lambda_{23\bar6 \bar7} \to 0$, $\lambda_{67\bar4 \bar5} \to 0$ and $\lambda_{23\bar4 \bar5} \to 0$. These threshold singularities can also occur simultaneously. The matching of threshold singularities happens pairwise between the two-loop phase-space residues $\ad{4,f}(123)$ and $\ad{4,f}(145)$, and the phase-space residue corresponding to the squared one-loop amplitude, $\ad{4,f}(167)$. The index $f=\Phi, \gamma^*$ denotes the specific decay process. In particular: 
\bea
&& \res{\ad{4,f}(167)\, \ps{\bar1 67} + \ad{4,f}(123) \ps{\bar1 23} , \lambda_{23\bar6 \bar7}} = 0~, \nn \\ 
&& \res{\ad{4,f}(167) \, \ps{\bar1 67}+ \ad{4,f}(145) \, \ps{\bar1 45}, \lambda_{67\bar4 \bar5}} = 0~, \nn \\ 
&& \res{\ad{4,f}(123) \, \ps{\bar1 23} + \ad{4,f}(145) \, \ps{\bar1 45}, \lambda_{23\bar4 \bar5}} = 0~, \nn \\ && 
\label{eq:double_threshold}
\eea
where the function $\ps{\bar i_1 i_2\cdots i_n} = 2\pi \delta\left( \lambda_{i_2\cdots i_n} - \lambda_{i_1}\right)$ implements energy conservation.
This matching of threshold singularities holds for the decays $\Phi\to \phi\phi (\phi \phi)$ and $\gamma^* \to q \bar q (gg)$, where the two-loop and squared one-loop phase-space residues correspond to the same final states. There is also a matching of threshold singularities for contributions with three particles in the final state, 
\bea
&& \res{\ad{4,f}(1278) \, \ps{\bar1 278} + \ad{4,f}(1479) \, \ps{\bar1 479}  
, \lambda_{28 \bar4 \bar 9}} \nn \\ && = 0~, \nn \\
&& \res{\ad{4,f}(1368) \, \ps{\bar1 368} + \ad{4,f}(1569) \, \ps{\bar1 569}  
, \lambda_{38 \bar5 \bar 9}} \nn \\ && = 0~. 
\eea
All these pairwise matchings of threshold singularities are encompassed by the proof presented in Ref.~\cite{Ramirez-Uribe:2024rjg}, which is valid to all perturbative orders. 

We may also consider the simultaneous occurrence of two or more thresholds. In such cases, the matching relations may depend on the specific loop topology. For example, for the one-loop squared and two-loop components in Fig~\ref{fig:nnlo} with $f=\Phi, \gamma^*$, we find
\bea
&& {\rm Res} \bigg(
{\rm Res} \bigg(
\ad{4,f}(167) \, \ps{\bar1 67} + \frac{1}{2} \ad{4,f}(123) \, \ps{\bar1 23} \nn \\ &&
+ \frac{1}{2} \ad{4,f}(145) \, \ps{\bar1 45}, 
\lambda_{23 \bar6 \bar 7} \bigg),  \lambda_{45 \bar6 \bar 7} \bigg) = 0~. 
\eea
This relation links all configurations where all particles are on the mass shell, with the exception of particles $8$ and $9$.

\section{Flavored constrained thresholds}

The decay $H\to \gamma\gamma$ is a special case because the leading order (LO) amplitude arises at one loop, and the corresponding decay rate is therefore given by the squared modulus of this one-loop amplitude. The choice of the specific $\gamma\gamma$ final state from the four-loop vacuum amplitude in Fig.~\ref{fig:nnlo}~(right) limits the phase-space contributions to $\ad{4,H}(167)$. For loop contributions with top quarks, threshold singularities are absent because of kinematics. Threshold singularities emerge only for lighter quarks. In that case, the matching of the threshold singularities cannot occur between different phase-space residues.

\begin{figure}[t]
\begin{center}
\includegraphics[width=.9\linewidth]{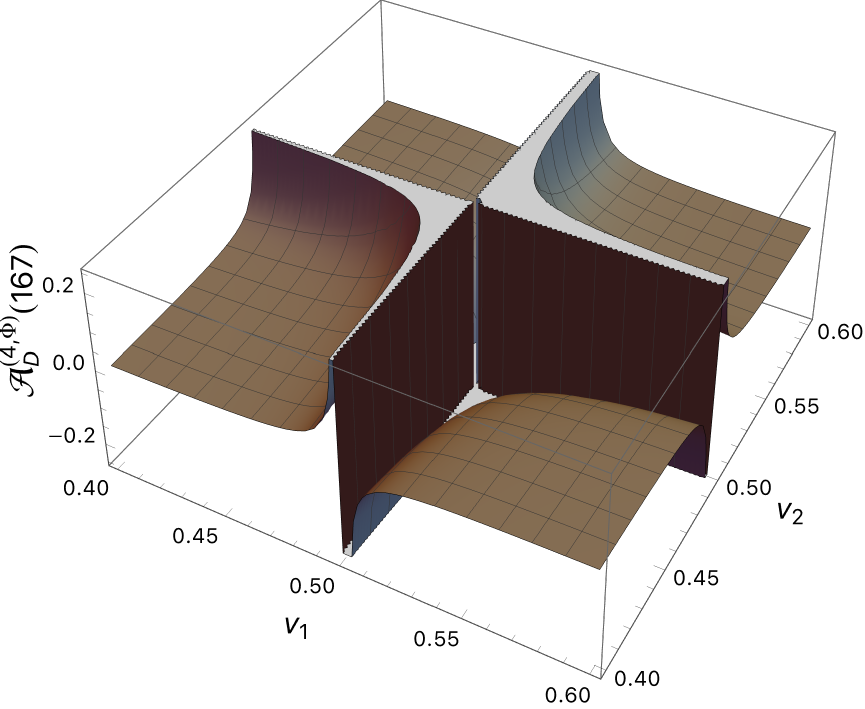}
\includegraphics[width=.9\linewidth]{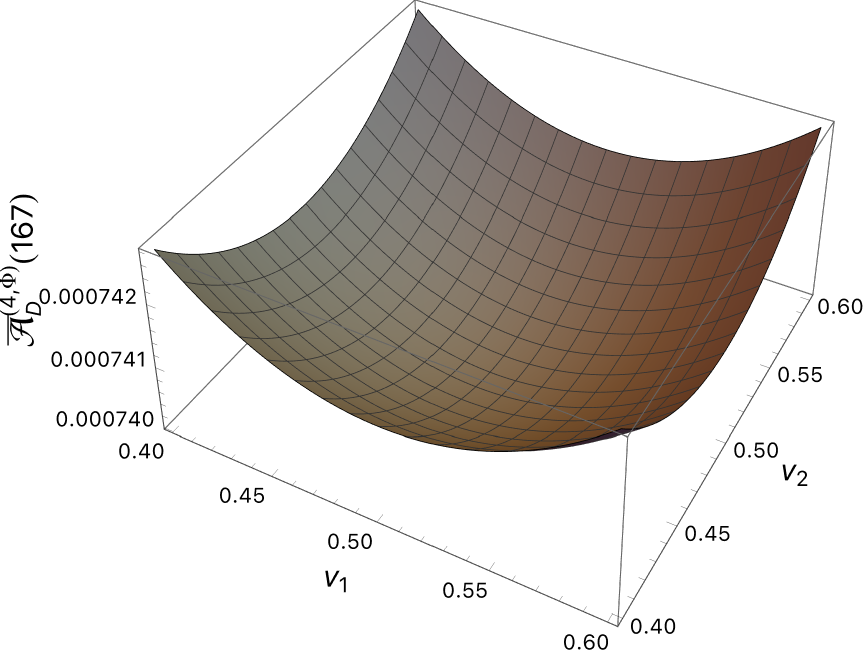}
\caption{Bare LTD integrand $\ad{4,\Phi}(167) \ps{\bar1 67}$, in which the threshold singularities are clearly visible (top), and after angular averaging resulting in a local cancellation (bottom). \label{fig:threshold}}
\end{center}
\end{figure}

To illustrate our solution to provide a threshold flat integrand representation, it is sufficient to consider the scalar vacuum amplitude of Fig.~\ref{fig:nnlo}~(left), where particles $6$ and $7$ are massless, particle $1$ is very heavy, and the rest share the same mass $m$. The bare decay rate is 
\beq
\Gamma^{(1,1)}_{\Phi\to \phi \phi} = 8\pi^2 s^2 \int_{\lb_1\lb_2\lb_4} \ad{4,\Phi}(167) \ps{\bar1 67}~.
\label{eq:decayratephi}
\eeq
The overall prefactor in \Eq{eq:decayratephi} has been chosen so as to define a bare decay rate which, although unphysical, is scaleless and normalized such that the integrated analytic result is
\beq
\Gamma^{(1,1)}_{\Phi\to \phi \phi} = {\rm Re} \left( {\cal A}^{(1)}_{\Phi\to \phi \phi} \left({\cal A}^{(1)}_{\Phi\to \phi \phi} \right)^\dagger \right)~, 
\eeq
where
\beq
{\cal A}^{(1)}_{\Phi\to \phi \phi} =
\frac{1}{32\pi^2} \lnn{\frac{\beta-1}{\beta+1}}~, 
\label{eq:analytic}
\eeq
with $\beta=\sqrt{1-4m^2/(s+\ii)}$, represent the scaleless one-loop amplitude. The numerical integration of the scalar one-loop amplitude in LTD by using a quantum computing approach has been analysed in Ref.~\cite{deLejarza:2024pgk}. The one-loop amplitude for $H\to \gamma\gamma$ has been calculated in LTD in Ref.~\cite{Driencourt-Mangin:2017gop} at LO, and in Ref.~\cite{Driencourt-Mangin:2019aix} at NLO.

The integrals in \Eq{eq:decayratephi} over the three-momenta $\lb_1$ and $\lb_2$ are mutually independent, and the energy conservation factor $\ps{167}$ fixes the modulus of $\lb_4$. Consequently, the integrand depends on the magnitudes of two three-momenta and on two angular variables. In the absence of a second phase-space residue, we implement an angular average over the angular variables. As shown in Fig.~\ref{fig:threshold}, this procedure effectively flattens the integrand across the threshold singularities by locally cancelling their divergent contributions. Fig.~\ref{fig:threshold} (top) presents the bare integrand as a function of the angular variables, $v_i = (1-\cos\theta_i)/2$, for arbitrary values of $|\lb_1|$ and $|\lb_2|$, in which the emergence of threshold singularities is obvious. Fig.~\ref{fig:threshold} (bottom) is the result of averaging over $v_i$ and $1-v_i$. This angular averaging induces a local cancellation of threshold singularities analogous to that illustrated in Ref.~\cite{LTD:2024yrb}, where the cancellation occurs between distinct phase-space residues, thereby yielding a well-behaved integrand suitable for numerical integration.

\begin{figure}[t]
\begin{center}
\includegraphics[width=.9\linewidth]{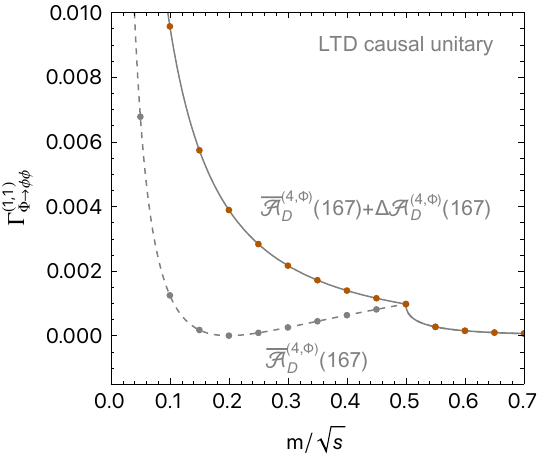}
\caption{Integrated decay rate for the process $\Phi\to \phi\phi$, with massless final-state scalars, as a function of the mass of the internal scalars. Gray dots represent the product of the real parts of the one-loop amplitudes. Orange dots are the full result, including the square of the imaginary parts. Dashed and solid lines are the corresponding analytic predictions.\label{fig:decayrate}}
\end{center}
\end{figure}

However, angular averaging alone does not provide the complete result, as it does not account for the configuration in which both threshold singularities occur simultaneously. Angular averaging provides an efficient procedure for a local removing of the imaginary contributions arising when only one of the two threshold singularities is active. Since we are evaluating the squared modulus of an amplitude, these imaginary contributions must cancel in the integrated result, although they remain present in the bare integrand. The missing contribution, corresponding to the square of the imaginary part, is obtained from the double residue
\bea
&& \Delta \ad{4,\Phi}(167) = \pi^2 \nn \\ && \quad \times \res{\res{\ad{4,\Phi}(167),\lambda_{23\bar6 \bar7}},\lambda_{45\bar6 \bar7}}~.
\eea
Therefore, our final expression of the decay rate is 
\bea
&& \Gamma^{(1,1)}_{\Phi\to \phi\phi} = 8\pi^2 s^2 \int_{\lb_1\lb_2\lb_4} \left( \overline{\cal A}_{\rm D}^{(4,\Phi)}(167) \right. \nn \\ && \qquad \left. + \Delta \ad{4,\Phi}(167) \right)  \ps{\bar1 67}~,
\label{eq:finaldecay}
\eea
where $\overline{\cal A}_{\rm D}^{(4,\Phi)}(167)$ is the angular average of the phase-space residue $\ad{4,\Phi}(167)$. The integrand in \Eq{eq:finaldecay} is well-behaved in all the integration domain. Results on the integrated decay rate are presented in Fig.~\ref{fig:decayrate}, and agree perfectly with the analytic predictions obtained from \Eq{eq:analytic} at the $10^{-4}$ level even using \texttt{NIntegrate} in Wolfram Mathematica. A more reliable assessment of the numerical statistical uncertainties could be obtained using VEGAS~\cite{Lepage:1977sw,Hahn:2004fe} or QAIS~\cite{Pyretzidis:2025stx}, which is not required for the present discussion, whose purpose is primarily illustrative.

\section{conclusions}

We have further examined the prescription introduced in Ref.~\cite{Ramirez-Uribe:2024rjg} for analytically continuing to negative values the on-shell energies of particles identified as initial-state particles in vacuum amplitudes, and have demonstrated unambiguously that it is well defined for all kinematical configurations.

We have expanded the discussion of the local matching of threshold singularities. For configurations where the threshold matching cannot be achieved between distinct phase-space residues, we have introduced angular averaging as an efficient procedure for constructing an LTD integrand representation in which these singularities are locally cancelled.

{\it Acknowledgments:} This work is supported by the Spanish Government and ERDF/EU - Agencia Estatal de Investigaci\'on MCIN/AEI/10.13039/501100011033,  Grants No. PID2023-146220NB-I00, No. EUR2025-164820, and No. CEX2023-001292-S. SRU acknowledges support from SECIHTI through the Sistema Nacional de Investigadoras e Investigadores (SNII). PKD is supported by the Swiss National Science Foundation (SNSF) under contracts 200020$\_$219367.

\bibliographystyle{JHEP}
\bibliography{2026_analytic_threshold}

\end{document}